\documentclass[%
 reprint,
 amsmath,amssymb,
 aps,
]{revtex4-2}

\usepackage{graphicx}
\usepackage{dcolumn}
\usepackage{bm}
\usepackage{amsthm}
\usepackage{slashed}
\usepackage{amsmath}
\usepackage{amssymb}
\usepackage{amsfonts}
\usepackage{mathrsfs}
\usepackage{mathtools}
\usepackage{xcolor}

\def\kbar{{\mathchar'26\mkern-9muk}}
\def\diff{\textrm{d}}

\def\dj{d\kern-0.4em\char"16\kern-0.1em}
\def\Dj{\mbox{\raise0.3ex\hbox{-}\kern-0.4em D}}

\begin{document}

\title{Rotating frames from quantum deformed spacetime}

\author{Du\v{s}an \Dj or\dj evi\'{c}}
\email{dusan.djordjevic@ff.bg.ac.rs}
\author{Dragoljub Go\v{c}anin}
\email{dragoljub.gocanin@ff.bg.ac.rs}
\affiliation{%
Faculty of Physics, University of Belgrade\\ Studentski Trg 12-16, 11000 Belgrade, Serbia
}%

\begin{abstract}

In a sense of deformation quantization, noncommutative (NC) geometry introduces a quantum structure of spacetime. Using the twist-deformation formalism, we show that the dynamical effects of spacetime noncommutativity can amount to a transition to a rotating frame of reference. In particular, we study the dynamics of charged matter (scalars and spinors) on the curved background of Melvin's electric universe in the framework of NC gauge field theory. Melvin's electric/magnetic universe is an exact sourceless solution of the Einstein-Maxwell field equations that is both static and axially symmetric, and it represents a parallel bundle of self-gravitating electric/magnetic flux. Due to its axial symmetry, it allows for a special kind of Killing twist that does not affect the coupling of the matter fields to the background metric. Focusing on the perturbative NC equations of motion for charged scalars and Dirac spinors coupled to Melvin's electric background, we show that they have the same form as the corresponding classical (undeformed) equations of motion coupled to the same geometric background but in a uniformly rotating frame whose angular velocity is determined by the NC scale and the electric charge of the matter field. In principle, this NC rotation effect can be experimentally tested by setting up the Sagnac interferometry apparatus to measure the Sagnac phase shift for charged versus electrically neutral particles, thus placing a bound on the scale of spacetime noncommutativity.      
\end{abstract}

\maketitle

\section{Introduction}

The application of noncommutative (NC) geometry in gravitational physics is based on the expectation that the classical description of spacetime as a smooth manifold breaks down at small enough length scales, closely resembling the way in which Heisenberg's uncertainty relations enforce the NC geometry structure on classical phase space.
In a more abstract setting, an NC spacetime is described by coordinate operators $\hat{x}^{\mu}$ that satisfy some non-trivial commutation relations, the simplest and the most common of which are the canonical, or $\theta$-constant commutation relations, $[\hat{x}^{\mu},\hat{x}^{\nu}]=i\theta^{\mu\nu}$,
where the NC deformation parameters $\theta^{\mu\nu}$ comprise a constant antisymmetric matrix, $\theta^{\mu\nu}=-\theta^{\nu\mu}$, characterizing the fundamental noncommutativity scale. The impossibility of sharply distinguishing individual spacetime events (points) is encoded by spacetime uncertainty relations  $\Delta\hat{x}^{\mu}\Delta\hat{x}^{\nu}\geq\frac{1}{2}\vert\theta^{\mu\nu}\vert$.
In analogy with the method of deformation quantization of classical phase space, we can implement the canonical spacetime noncommutativity by keeping the underlying commutative (i.e. undeformed, classical) structure of spacetime and use the (invertible) twist operator 
\begin{equation}
 \mathcal{F}=\exp\left(-\frac{i}{2}\theta^{\mu\nu}\partial_{\mu}\otimes\partial_{\nu}\right)   
\end{equation}
to deform instead the algebra of functions of ordinary commuting coordinates, $x^{\mu}$, by promoting ordinary pointwise multiplication to the noncommutative, but associative, Moyal-Weyl-Groenewold (MWG) $\star$-product, 
\begin{align}\label{Moyal}
(f\star g)(x)&=\mu\left(\mathcal{F}^{-1}(f\otimes g)\right)\\
&=f(x)g(x)+\frac{i}{2}\theta^{\mu\nu}\partial_{\mu}f(x)\partial_{\nu}g(x)+\mathcal{O}(\theta^{2}),\nonumber
\end{align}
where $\mu\left(f\otimes g\right)=f\cdot g$ is the multiplication map. 
Note that, when applied on coordinate functions themselves (treated as scalar fields), it gives us the $\theta$-constant $\star$-commutator relations,
$[x^{\mu},x^{\nu}]_{\star}=x^{\mu}\star x^{\nu}-x^{\nu}\star x^{\mu}=i\theta^{\mu\nu}$.  

By definition, the MWG $\star$-product is adapted to a particular system of coordinates, at least in some patch of spacetime. However, we can generalize the twist operator in a coordinate-free manner. For a set $\{X_I\, \vert\, I=1,\dots,s\leq D\}$ of \emph{mutually commuting} vector fields (note that the number of vector fields $s$ need not be equal to the number of spacetime dimensions $D$, in which case $\{X_{I}\}$ span, at each point, a subspace of a $D$-dimensional tangent space) and a constant antysymmetric matrix $\theta^{IJ}$, an \textit{abelian} Drinfeld twist is given by
\begin{equation}
  \mathcal{F}
  =\exp\left(-\frac{i}{2}\theta^{IJ}X_I\otimes X_J\right),
\end{equation}
with the associated $\star$-product 
\begin{align}\label{star}
 (f\star g)(x) &= \mu\left(\mathcal{F}^{-1}
    (f\otimes g)\right)\nonumber\\
    &=f(x)g(x)+\frac{i}{2}\theta^{IJ}X_{I}[f]X_{J}[g]+\mathcal{O}(\theta^{2}),
\end{align}
where $X_{I}[f]=X^{\mu}_{I}\partial_{\mu}f$.
Since the vector fields \(X_I\) commute, the \(\star\)-product is \textit{associative}. Some more details on the twist formalism can be found in Appendix A. We want to emphasize that the system of commuting vector fields $\{X_{I}\}$ that generate the twist operator is an \textit{independent} structure on the spacetime manifold that provides a generally covariant formulation of an NC theory. Of course, we can always choose (at least locally) a system of coordinates $x^{\mu}$ that is adapted to $\{X_{I}\}$ by setting $X_{I}=\partial_{\mu}$ so that (\ref{star}) reduces to the MWG $\star$-product (\ref{Moyal}) and the $\star$-commutator becomes $[x^{\mu},x^{\nu}]_{\star}=i\theta^{IJ}\delta^{\mu}_{I}\delta^{\nu}_{J}=i\theta^{\mu\nu}=const$. In some other system of coordinates, unrelated to $\{X_{I}\}$, the $\star$-commutator, in general, would not be constant. In other words, the choice of $\{X_{I}\}$ determines in which coordinates the \textit{constant} noncommutativity is to be realized. A comprehensive review of the concepts and methods of an NC field theory/gravity, including the twist formalism, can be found in \cite{NC_book}.

Although essential for defining a $\star$-product, the vector fields that generate the twist make an NC theory explicitly background dependent. This feature suggests that there might exist some underlying dynamical system that naturally brings about these vector fields in the low-energy limit. If so, there should also be some kind of physical criterion to determine what set of vector fields we should choose in a particular situation. However, in the absence of such criteria, it seems reasonable to motivate our choice by symmetry, namely the \textit{isometries} of spacetime. Let $g_{\mu\nu}(x)\diff x^{\mu}\diff x^{\nu}$ be the metric tensor in some coordinate system $x^{\mu}$. A Killing vector field, $K$, satisfies the Killing equation $\mathcal{L}_{K}g=0$, where
\begin{equation}
(\mathcal{L}_{K}g)_{\mu\nu}=
K^{\alpha}\partial_{\alpha}g_{\mu\nu}
+g_{\alpha\nu}\partial_{\mu
}K^\alpha+g_{\mu\alpha}\partial_{\nu
}K^\alpha.
\end{equation}
Taking a set of Killing vector fields $\{K_{I}\}$ of a given spacetime as generators of an abelian Drinfled twist ensures that while the algebraic structure of the theory gets NC deformed, the coupling with the metric components and their functions (connection, curvature) remain unchanged. Namely, consider a product $g_{\mu\nu}(x)\cdot(\cdots)$ of the metric component with something. If we were to deform the ordinary product by a generic twist (not a Killing one), up to the first order in $\theta$, we would get 
\begin{equation}
g_{\mu\nu}\star(\cdots)=g_{\mu\nu}\cdot(\cdots)+
\frac{i}{2}\theta^{IJ}X_{I}[g_{\mu\nu}]\mathcal{L}_{X_{J}}(\cdots),
\end{equation}
where $\mathcal{L}_{X_{J}}$ is the Lie derivative along $X_{J}$ and $X_{I}[g_{\mu\nu}]=X_{I}^{\alpha}\partial_{\alpha}g_{\mu\nu}$, which generally is not zero. Thus, for a general abelian twist, other fields are coupled to geometry in an NC way via the $\star$-product. Now, if we take the Killing vector fields $\{K_{I}\}$ as generators of an abelian twist, and make a transition to the coordinate system $y^{\alpha}$ satisfying $K_{I}=\delta^{\alpha}_{I}\tfrac{\partial}{\partial y^{\alpha}}\equiv \delta^{\alpha}_{I}\partial_{\alpha}$ (the number of Killing vector fields may be smaller than the number of spacetime dimensions) then we have $(K_{I})^{\alpha}=\delta^{\alpha}_{I}$, and the Killing equation yields $\partial_{\alpha}g_{\mu\nu}(y)=0$. For a non-Killing twist, even if we adapt our coordinates to the twist-generating vector fields, the Killing condition demands $X_{I}[g_{\mu\nu}]\neq0$. Therefore, only by choosing to work with a Killing twist, we can ensure that the coupling with the metric remains commutative as in the classical theory. 
 
While it leaves the coupling with spacetime geometry unchanged, a Killing twist can result in NC deformed equations of motion for matter fields. Moreover, in \cite{K1, K2} it was shown that, in certain situations, these NC equations of motion for matter fields, obtained by a Killing twist, can be rewritten (at least perturbativelly) as their classical counterparts but for an \textit{effective} metric representation that takes into account the NC corrections - the so-called \textit{NC duality}.  Schematically, for a massless scalar field $f$, one would obtain an NC field equation of the form
\begin{equation}
(\Box_{g}+\mathcal{O}(\theta))f=0, 
\end{equation}
where $g_{\mu\nu}$ is the original representation of the spacetime metric and $\mathcal{O}(\theta)$ contains the NC corrections. It turns out that one can rewrite the previous equation in the classical form
\begin{equation}
\Box_{G}f=0, 
\end{equation}
where $G_{\mu\nu}=g_{\mu\nu}+\mathcal{O}(\theta)$ are effective NC metric components. The effective metric representation can be very different from the original one, raising a question about its physical interpretation. Typically, it turns out that the two are related by a coordinate transformation, thus describing the same underlying geometry. However, due to the coupling with matter fields, their difference may become observable and thus physically meaningful. In this respect, the NC duality was studied in \cite{T1, T2, T3, T4}. We have pointed out earlier that the choice of vector fields that generate the twist determines the coordinate system in which the noncommutativity is realized. One can certainly come up with some convenient choice of vector fields that does lead to an effective description that might involve some interesting new dynamics. However, we hold that this ad hoc construction of the twist operator designed to make some NC effect apparent is not entirely appropriate. Therefore, we advocate that the choice of the vector fields should be based on isometries and that the relevant effective NC description can come only through a Killing twist.

For that matter, in this paper, we consider the dynamics of electrically charged matter fields (both scalars and Dirac spinors) in Melvin's \textit{electric} universe \cite{Mel1, Mel2, Mel3, Mel4} and show, by using a suitable Killing twist, that one can pass to an effective description in which NC corrections, at least up to the first order in $\theta$, can be accounted for by a simple transformation to a rotating frame. In other words, by introducing spacetime noncommutativity, as far as the dynamics of charged matter is concerned, one effectively makes a transition to a (uniformly) rotating frame of reference, with the angular velocity determined by the scale of noncommutativuty and the charge of the matter field. In this way, the new NC interaction terms that come from the twist deformation receive a purely kinematical interpretation, as if they were some kind of ``inertial forces''. We point out that the effect appears only if the matter is electrically charged; in the case of electrically neutral fields the dynamics is not altered. This selective sensitivity to the purported NC structure of spacetime is the main reason for regarding the effect of ``NC rotation'' as genuinely physical and potentially observable, even though it pertains to a particular model that is not entirely realistic.  

This naturally leads to the consideration of the Sagnac effect in the context of NC geometry. 
The study of the Sagnac effect, originally observed in 1913 by Georges Sagnac using a rotating optical interferometer \cite{Sagnac} has a long tradition. For a historic perspective on the subject of interferometric tests of rotational effects, see \cite{Sagnac1} and the references therein. The Sagnac effect is a fundamental physical phenomenon that connects rotating frames, relativistic geometry, and phase coherence. It manifests as a measurable time delay (or phase shift) between two waves (of light or two quantum wave packets generated by a beam splitter) traveling in opposite directions along a closed loop in a rotating frame. Upon recombination, an interference pattern shift is detected, the magnitude of which is proportional to the angular velocity of the frame and the area enclosed by the loop. While the original intent of Sagnac's experiment was to argue against Einstein's special theory of relativity in favor of a stationary ether, it is now well-understood that the Sagnac effect arises naturally within special relativity and is fully compatible with its geometric structure.

In special relativity, the origin of the Sagnac effect is best understood through the non-existence of global simultaneity surfaces for rotating observers. In rotating frames, time cannot be defined globally in a consistent way using Einstein synchronization, due to the presence of non-zero vorticity in the congruence of observer worldlines.
Although each observer has a well-defined local notion of simultaneity, these local slices cannot be consistently patched together into a global hypersurface. This failure of global clock synchronization (the so-called synchronization gap) directly leads to an observable phase shift in interferometric experiments. 

Special relativity predicts the following proper‐time difference
(as measured by a clock at rest in the starting/ending point on the
rotating ring, i.e. the beam splitter point) between counter‐propagating beams in a ring interferometer:
\begin{equation}\label{SagnacSTR}
  \Delta \tau
  = \frac{4\pi R^2\Omega}{c^2\sqrt{1 - \frac{\Omega^2 R^2}{c^2}}}
    \approx \frac{4\pi R^2\Omega}{c^2} ,
\end{equation}
where $R$ is the radius of the ring and $\Omega$ its (constant) angular speed \cite{STR}; the nonrelativistic approximation holds for $\tfrac{\Omega R}{c}\ll1$.
The previous formula highlights the universality of the Sagnac effect. Namely, the time \textit{difference} (and thus the phase shift) is independent of the nature of the interfering objects (such as classical electromagnetic and acoustic waves or quantum particles like electrons and neutrons) counter-propagating along a closed path in a rotating interferometer, with
the same (in absolute value) velocity with respect to the interferometer; it depends only on the geometry of the path and the rotation rate.  

To generalize this result to curved spacetime, we follow the systematic approach developed by Rizzi and Ruggiero \cite{Sagnac2}. Their method emphasizes the geometric and topological character of the Sagnac effect by interpreting it as a holonomy in the space of simultaneity slices, enforcing the analogy with the Aharonov-Bohm phase shift in gauge theories given by
\begin{equation}
 \Delta \phi =\frac{q}{c}\oint_{C}\boldsymbol{A}\cdot\diff\boldsymbol{l}=\frac{q}{c} \int_S \boldsymbol{B} \cdot 
\diff\boldsymbol{S}.   
\end{equation} 
The method from \cite{Sagnac2} amounts to a fully relativistic generalization of the original Sakurai's replacement trick \cite{Sakurai}, $\tfrac{q}{c} \mathbf{B} \rightarrow 2 m \mathbf{\Omega}$, based on a formal analogy between the Lorentz force and the Coriolis force, which leads to the correct nonrelativistic approximation for the Sagnac phase shift (\ref{SagnacSTR}). By applying Cattaneo’s splitting formalism \cite{Cat1, Cat2, Cat3, Cat4, Cat5}, the dynamics of massive or massless
particles, relative to a given time-like congruence of unit vector fields $u^{\mu}(x)$ (4-velocity field satisfying $g_{\mu\nu}u^{\mu}u^{\nu}=-1$) can
be described in terms of a ``gravito-electromagnetic'' potentials $\Phi^{G}=-c^{2}u^{0}$ and $A_{i}^{G}=c^{2}\tfrac{u_{i}}{u_{0}}$, modifying the Sakurai's replacement by 
$\tfrac{q}{c}\mathbf{B} \to \tfrac{m u_0}{c}   \mathbf{B}^G$ (where $\mathbf{B}^G$ is related to the vorticity of the congruence), yielding the Sagnac phase shift as a gravito-magnetic Aharonov-Bohm shift,
\begin{equation}
\Delta \phi = \frac{2m u_0}{\hbar c} \oint_C \boldsymbol{A}^G \cdot \diff \boldsymbol{l} = -\frac{2m\sqrt{-g_{00}}}{\hbar c} \int_S \boldsymbol{B}^G \cdot \diff\boldsymbol{S}.   
\end{equation}
This formula consistently reproduces the flat spacetime result and directly generalizes 
the Sagnac effect to curved space-time, which is of particular interest for this paper.

This geometric insight regarding the Sagnac phase finds a natural continuation in quantum theory, where phase coherence and topological effects play an important role. Quantum versions of the Sagnac experiment have confirmed the existence of a rotation-induced phase shift proportional to the enclosed area and angular velocity, independent of the internal structure of the particle \cite{SQM1, SQM2, SQM3, SQM4}. Such interferometric setups are the basis for rotation-sensitive quantum technologies.
In addition to their technological significance, these quantum Sagnac experiments have theoretical implications for our understanding of spacetime. The emergence of quantum holonomies in non-inertial settings has led to explorations of how entanglement, superposition, and locality behave in rotating frames \cite{QMrot1, QMrot2, QMrot3}. 
These subtleties motivate a deeper analysis of the Sagnac effect in quantum regimes where the geometric structure of spacetime may depart from classical notions.

In NC geometry, where the underlying spacetime loses its point-like structure, and the concepts of trajectory, simultaneity, and locality become fuzzier, interference phenomena, being sensitive to global geometric and topological features, are expected to be particularly revealing of these new structures. Several works have explored the implications of NC geometry for quantum phase effects. Bellucci and Nersessian \cite{Bellucci} analyzed the behavior of quantum phases in NC quantum mechanics on curved (pseudo)spherical backgrounds, showing that the NC deformation alters the structure of Berry phases. Chaichian et al. \cite{Chaichian} showed that even the hydrogen atom spectrum and Lamb shift receive corrections due to noncommutativity, leading to rotational anisotropies. Mirza and Zarei \cite{Mirza} demonstrated that the Aharonov-Casher effect, which is closely related to the Sagnac effect in its dependence on path topology and spin structure, also acquires NC corrections. NC models emerge in several approaches to quantum gravity, most famously in string theory \cite{SW}. The above mentioned findings, however, suggest that the Sagnac effect, especially in quantum mechanical settings, could be modified by noncommutativity, offering a potential low-energy window into Planck-scale physics.

In this work, we explore the connection between rotating reference frames and NC geometry in terms of a Killing twist deformation of Melvin's electric universe. We propose a setup for determining whether interferometric phase shifts in rotating frames can serve as sensitive probes of spacetime noncommutativity, and what such measurements can reveal about the small-scale structure of spacetime. In the following section, we analyze the dynamics, both classical and noncommutative, of an electrically charged scalar field on Melvin's electric background. In Section III we present the NC Sagnac effect. Discussion and outlook are left for the final section. 



\section{The model}

Consider the action for a massive complex scalar field $f$ (of mass $m$ and charge $q$) in the background of a $U(1)$ gauge field $A_{\mu}$ and a (curved) four-dimensional spacetime with metric $g_{\mu\nu}$,
\begin{align}
& \int \diff^4 x \sqrt{|g|}\left(g^{\mu\nu} D_\mu f^* D_\nu f + m^{2} f^* f\right),
 \end{align}
where $D_{\mu}f=\partial_{\mu}f-iqA_{\mu}f$ is the $U(1)$ covariant derivative. We set $\hbar=c=G=1$. The gauge field strength is $F_{\mu\nu}=\partial_{\mu}A_{\nu}-\partial_{\nu}A_{\mu}$. Varying with respect to $f^*$ we get the equation of motion for the scalar field,
\begin{equation}\label{EoM}
g^{\mu\nu}\bigl(\partial_{\mu}-\Gamma_{\mu}-iqA_{\mu}\bigr)D_{\nu}f
-m^{2}f=0.
\end{equation}
The Christoffel symbols $\Gamma^{\lambda}_{\mu\nu}$ appear since the background metric $g_{\mu\nu}$ is not flat. As a geometric background, we consider a
very special solution of the classical Einstein-Maxwell equations known as Melvin's electric/magnetic universe \cite{Mel1, Mel2, Mel3, Mel4}. This geon is a static, sourceless, nonsingular, axially symmetric configuration of electro-magnetic field contained by the curvature of spacetime generated by their own energy density. It can be 
represented as a parallel bundle of electric/magnetic flux held together by its own gravitational pull. Although the magnetic case is more involved in the literature due its cosmological applications, here we will focus on Melvin's electric universe. The gauge potential $A_{0}=Ez$ describes an electrostatic field in the $z$-direction. In stationary cylindrical coordinates $(t,\rho, z,\varphi)$ the metric of Melvin's electric universe reads
\begin{equation}
ds^{2}=\lambda^{2}(\rho)[-\diff t^{2}+\diff\rho^{2}+\diff z^{2}]+\frac{\rho^{2}}{\lambda^{2}(\rho)}\diff \varphi^{2},  
\end{equation}
with $\lambda(\rho)=1+\tfrac{1}{4}E^{2}\rho^{2}$ \cite{Mel4}. The non-zero components of the connection $\Gamma$ are the following:
\begin{align}
\Gamma_{00}^{1}&=\Gamma_{11}^{1}=\frac{1}{\lambda(\rho)}\partial_{\rho}\lambda(\rho),\nonumber\\
\Gamma_{22}^{1}&=-\frac{1}{\lambda}\partial_{\rho}\lambda(\rho),\nonumber\\
\Gamma_{33}^{1}&=-\frac{1}{2}\frac{1}{\lambda^{2}(\rho)}\partial_{\rho}\left(\frac{\rho^{2}}{\lambda^{2}(\rho)}\right).
\end{align}
For future reference, note that if we transform to a non-inertial reference frame rotating around the symmetry axis with uniform angular velocity $\Omega>0$, its coordinates $(t',\rho',z',\varphi')$ are related to the original ones by $t'=t$, $\rho'=\rho$, $z'=z$ and $\varphi'=\varphi-\Omega t$, and Melvin's metric takes the following form,
\begin{align}\label{MelvinRot}
ds^{2}=&\lambda(\rho)[-c^{2}\diff t^{2}+\diff\rho^{2}+\diff z^{2}]+\frac{\rho^{2}}{\lambda^{2}(\rho)}(\diff\varphi'+\Omega \diff t)^{2}\nonumber\\
=&\left(-c^{2}\lambda^{2}(\rho)+\frac{\rho^{2}\Omega^{2}}{\lambda^{2}(\rho)}\right)\diff t^{2}+\lambda^{2}(\rho)[\diff\rho^{2}+\diff z^{2}]\nonumber\\
&+\frac{\rho^{2}}{\lambda^{2}(\rho)}\diff\varphi'^{2}+\frac{\rho^{2}\Omega}{\lambda^{2}(\rho)}(\diff\varphi' \diff t+\diff t\diff\varphi').
\end{align}
Evaluating (\ref{EoM}) on Melvin's electric background yields
\begin{equation}
\left[\Box_{g}-m^{2}+\frac{2iqEz}{\lambda^{2}(\rho)}\partial_{t}+\frac{1}{\rho\lambda^{2}(\rho)}\partial_{\rho}+\frac{q^{2}E^{2}z^{2}}{\lambda^{2}(\rho)}\right]f=0,        
\end{equation}
where $\Box_{g}=g^{\mu\nu}\partial_{\mu}\partial_{\nu}$. Our first goal is to find the perturbative NC correction to this classical equation of motion. 

To construct an NC gauge
field theory action from a given classical (commutative) one, we use the technology of twist deformations. 
An NC action has the same form as its classical counterpart, except that now the $\star$-product replaces the ordinary product. However, simply inserting the $\star$-product in the classical action is not enough to render its NC generalization; we have to take care of the symmetry. To maintain gauge invariance of the NC theory, one also has to introduce NC fields (we will denote them by a hat symbol) whose prime feature is that they change under $\star$-deformed gauge transformations in the same manner as classical fields change under ordinary gauge transformations. In our case, the NC $U(1)_{\star}$ action for a massive complex scalar field in a fixed (but arbitrary) spacetime background is
\begin{align}
& \int \diff^4 x \sqrt{|g|}\star\left(g^{\mu\nu} \star D_\mu \hat{f}^* \star D_\nu \hat{f} + m^{2} \hat{f}^* \star \hat{f}\right),
 \end{align}
where $D_{\mu}\hat{f}=\partial_{\mu}\hat{f}-iq\hat{A}_{\mu}\star\hat{f}$ is the $U(1)_{\star}$ covariant derivative and there are two NC fields, $\hat{f}$ and $\hat{A}_{\mu}$. Note that the metric is not promoted to an NC field because we only want to deform the algebra of $U$(1) gauge transformations while keeping the background geometry fixed. This ''semi-NC`` construction resembles the familiar semi-classical approach of quantum field theory in curved spacetime where we quantize the matter fields while the geometry remains classical.  

We take Melvin's electric universe as our classical background. Being static and axially symmetric, we have three commuting Killing vector fields at our disposal: $\partial_{t}$, $\partial_{z}$ and $\partial_{\varphi}$. Consider an (abelian) Killing twist generated by $X_1=\partial_z$ and $X_2=\partial_\varphi$,   
\begin{equation}
\mathcal{F}=e^{-i\frac{\kbar}{2}(\partial_z\otimes \partial_\varphi-\partial_\varphi\otimes \partial_z )}. 
\end{equation}
Note that $\theta^{\mu\nu}=\theta^{IJ}\delta^{\mu}_{I}\delta^{\nu}_{J}$ and so $\theta^{z\varphi}=-\theta^{\varphi z}\equiv\kbar$ with all other entries of $\theta^{\mu\nu}$ equal to zero. This type of twist, using the convention of \cite{Gubitosi, Fabiano}, is referred to as the $\lambda$-Minkowski twist. Of course, it was originally used to deform the Minkowski plane, hence the name.   

Being a Killing twist with respect to the background metric, the corresponding $\star$-product acts trivially on the metric and its functions, and we can simply omit it in the NC action, thus obtaining
\begin{align}
&\int \diff^4 x \sqrt{|g|}\left(g^{\mu\nu} D_\mu \hat{f}^* \star D_\nu \hat{f} + m^{2} \hat{f}^* \star \hat{f}\right).
 \end{align}
For an NC gauge parameter $\hat{\epsilon}$, the infinitesimal $U(1)_{\star}$ gauge variations of the fields are given by     
\begin{align}
\delta_{\hat{\epsilon}}\hat{f}&=i\hat{\epsilon}\star\hat{f},\\
\delta_{\hat{\epsilon}}\hat{A}_{\mu}&=-\partial_{\mu}\hat{\epsilon}+iq[\hat{\epsilon},\hat{A}_{\mu}]_{\star},\\
\delta_{\hat{\epsilon}}g_{\mu\nu}&=0.
\end{align}
The NC action is invariant under these NC deformed $U(1)_{\star}$ gauge transformations, by construction. 

The NC action can be organized as a perturbative expansion in powers of $\theta^{\mu\nu}$ using the 
Seiberg-Witten (SW) map that allows us to express the NC degrees of freedom in terms of the original, commutative ones \cite{SW}. We do this by imposing the following relation
\begin{equation}
\delta_{\hat{\epsilon}}\hat{A}_\mu(A)=\hat{A}_\mu(A+\delta_{\epsilon}A)-\hat{A}_\mu(A), \end{equation} 
where we assume that the NC gauge transformations are induced from the commutative ones. Using the commutative, $\delta_{\epsilon}A_{\mu}=-\partial_{\mu}\epsilon$, and the noncommutative, $\delta_{\hat{\epsilon}}\hat{A}_{\mu}=-\partial_{\mu}\hat{\epsilon}+iq[\hat{\epsilon},\hat{A}_{\mu}]_{\star}$, variation laws we can solve the previous differential equation perturbatively and derive the nonlinear SW map that represents the NC field $\hat{A}$ as a power series in $\theta^{\mu\nu}$ with coefficients built out of classical fields (likewise for $\hat{f}$), 
\begin{align}
\hat A_{\mu}
&= A_{\mu}
- \frac{1}{2}\theta^{\rho\sigma}A_{\rho}
    \left(\partial_{\sigma}A_{\mu} + F_{\sigma\mu}\right)+\mathcal{O}(\theta^{2}),\\
    \hat f 
&= f 
- \frac{1}{4}\,\theta^{\mu\nu}A_{\mu}
    \left(\partial_{\nu}f + D_{\nu}f\right)+\mathcal{O}(\theta^{2}).
\end{align}
\if
However, the fact that UEA is infinite-dimensional implies an infinite number of unwanted new degrees of freedom. 
Seiberg-Witten (SW) map allows us to redefine these new NC degrees of freedom in terms of the original classical ones \cite{SW}. The main idea of Seiberg and Witten was to show that these UEA-valued NC fields that transform under NC gauge transformations can be organized into a perturbation series in powers of $\theta$, with coefficients built out of fields from the commutative theory subjected to ordinary gauge transformation laws. 
The basic principle behind the SW construction is that NC gauge transformations are induced by the corresponding commutative ones,
\begin{equation}
\delta_{\hat{\epsilon}}\hat{A}(A)=\hat{A}(A+\delta_{\epsilon}A)-\hat{A}(A),    
\end{equation}
where the NC gauge field $\hat{A}$ is a function of the classical gauge field $A$. Using the commutative, $\delta_{\epsilon}A_{\mu}=-\partial_{\mu}\epsilon$, and the noncommutative, $\delta_{\hat{\epsilon}}\hat{A}_{\mu}=-\partial_{\mu}\hat{\epsilon}+iq[\hat{\epsilon},\hat{A}_{\mu}]_{\star}$, variation laws we can solve the previous differential equation perturbatively and derive the nonlinear SW map that represents the NC field $\hat{A}$ as a power series in $\theta$ with coefficients built out of classical fields (likewise for $\hat{f}$), 
\begin{align}
\hat A_{\mu}
&= A_{\mu}
- \frac{1}{2}\theta^{\rho\sigma}A_{\rho}
    \left(\partial_{\sigma}A_{\mu} + F_{\sigma\mu}\right)+\mathcal{O}(\theta^{2}),\\
    \hat f 
&= f 
- \frac{1}{4}\,\theta^{\mu\nu}A_{\mu}
    \left(\partial_{\nu}f + D_{\nu}f\right)+\mathcal{O}(\theta^{2}).
\end{align}
Furthermore, the SW map allows us to expand the NC action in powers of $\theta$ and ensures its invariance under ordinary gauge transformations, order by order. The leading order term ($\theta=0$) is the original classical action, while higher order terms represent NC corrections that can be interpreted as new interactions for classical fields. 
\fi
In our case, the NC action of order $\mathcal{O}(\kbar)$ is 
\begin{align}
&\int \diff^4x \sqrt{|g|} \Big[
    g^{\mu\nu} D_\mu f^* D_\nu f+m^{2}\vert f\vert^{2}-\frac{m^{2}}{4}\theta^{\alpha\beta}F_{\alpha\beta}\vert f\vert^{2}
      \nonumber\\
      &+ \frac{1}{2} \theta^{\alpha\beta} g^{\mu\nu}\Big(- \frac{1}{2}(D_\mu f)^* F_{\alpha\beta} D_\nu f\nonumber\\
    &+ (D_\mu f)^* F_{\alpha\nu} D_\beta f
    + (D_\beta f)^* F_{\alpha\mu} D_\nu f
\Big)
\Big].
\end{align}
The perturbative NC action is invariant under undeformed $U(1)$ gauge transformations. We see that the NC twist introduces new interaction terms for the original classical fields that will change the equation of motion for $f$. Up to first order in $\kbar$, the NC equation of motion for Melvin's electric background is 
\begin{align}
\Bigg[\Box_{g}-m^{2}&+\frac{2iqEz}{\lambda^{2}(\rho)}\partial_{t}+\frac{1}{\rho\lambda^{2}(\rho)}\partial_{\rho}\\
&+\frac{q^{2}E^{2}z^{2}}{\lambda^{2}(\rho)}
-\frac{qE\kbar}{\lambda^{2}(\rho)}\Big(\partial_{t}\partial_{\varphi}-iqEz\partial_{\varphi}\Big)\Bigg]f=0.\nonumber
\end{align}
It is a remarkable feature of this (perturbative) NC equation of motion that it can be rewritten as the classical equation of motion for the scalar field but in a different, effective NC metric representation $G_{\mu\nu}$ given by
\begin{align}\label{NCmetric}
\diff s^{2}=&G_{\mu\nu}(y)\diff y^{\mu}\diff y^{\nu}\nonumber\\
=&\lambda^{2}(\rho)(-\diff t^{2}+\diff \rho^{2}+\diff z^{2})+\frac{\rho^{2}}{\lambda^{2}(\rho)}\diff \varphi^{2}\nonumber\\
&-\frac{qE\kbar\rho^{2}}{2\lambda^{2}(\rho)}(\diff t\diff \varphi+\diff \varphi\diff t),     
\end{align}
as if the NC correction is absorbed by the effective metric - a sort of ''NC trade-off``. The noncommutativity appears in the off-diagonal components $G_{t\varphi}=G_{\varphi t}$ of the effective metric. Note, however, that the underlying geometry is not changed, only the coordinate system in which we represent the metric components. Up to a correction of order $\mathcal{O}\left(\theta^2\right)$, this metric can be written as (\ref{MelvinRot}) with an NC angular velocity determined by the NC parameter and the electric charge of the field,
\begin{equation}
\Omega_{\text{NC}}=-\frac{1}{2}qE\kbar.
\end{equation}
This implies that new NC couplings change the dynamics of complex scalar field in Melvin's electric universe in a way that is equivalent (at least perturbatively) to a transformation to a rotating frame of reference. 
It is important to note that electrically neutral fields do not pick up this NC effect, thus reinforcing the claim that this is not just an artifact of the formalism but a genuine physical effect. If we had two types of particles, ones that are charged and the others that are not, we should be able to see the experimental consequences of this ``NC rotation'', as the latter would not feel as if they were in the rotating frame, but the former ones would. In the following section, we will describe the setup for testing this NC effect based on Sagnac interferometry.  


But before we move on to the NC Sagnac effect, let us briefly extend the previous analysis by introducing the (positive) cosmological constant $\Lambda$ in Melvin's universe \cite{Zofka}. It leads us to the following metric ($\sigma$ is a constant parameter),
\begin{equation}
\diff s^2=-\diff t^2+\diff \rho^2+\diff z^2+\sigma^2\sin^2(\sqrt{2\Lambda}\rho)\diff \varphi^2.
\end{equation}
In the electric case, the gauge field is $A_{0}=\sqrt{\Lambda}z$ 
and the NC equation of motion for $f$ reads
\begin{align}
   \Bigg[\Box_{g_{\Lambda}}-m^{2}&+\sqrt{\Lambda } \left(\sqrt{2} \cot \left(\sqrt{2} \sqrt{\Lambda } \rho \right) \partial_\rho +2 i
   q z \partial_t\right)\\
   &+\Lambda  q^2 z^2   -q \kbar  \Big(\sqrt{\Lambda } \partial_t\partial_\varphi-i \Lambda  q z \partial_\varphi \Big)\Bigg]f=0.\nonumber
\end{align}
Again, we can rewrite the NC equation in terms of an effective metric given by
\begin{align}
\diff s^2=&-\diff t^2+\diff \rho^2+\diff z^2+\sigma^2\sin^2(\sqrt{2\Lambda}\rho)\diff \varphi^2\nonumber\\
&-\frac{1}{2}\kbar\sqrt{\Lambda}q\sigma^{2}\sin^{2}(\sqrt{2\Lambda}\rho)(\diff t\diff \varphi+\diff\varphi\diff t),  
\end{align}    
which describes the $\Lambda>0$ Melvin's electric universe from a uniformly rotating frame with $\Omega_{\text{NC}}=-\frac{1}{2}\kbar q\sqrt{\Lambda}$. This form of spacetime was considered recently in \cite{Oliveira}. Note that the scalar field in the magnetic Melvin universe in a noninertial rotating reference frame was recently analysed in \cite{Barbosa:2025bdc}, but without any reference to NC physics. 

Finally, we show in Appendix B that charged fermions exhibit the same behavior, with the same effective NC metric in both $\Lambda=0$ and $\Lambda>0$ case. However, if we ignore the spin-orbital coupling in the rotating frame, the scalar field will suffice.    

\section{Noncommutative Sagnac Effect}

Now that we have our main theoretical prediction set forth, namely that the dynamical response of electrically charged matter (as opposed to electrically neutral matter) to NC geometry can be (at least perturbatively) taken into account purely \textit{kinematically} by transforming to a rotating frame of reference, we want to discuss potential ways in which this effect of NC rotation could be observed. Of course, for that matter, one would have to generalize our analysis to some more realistic model of spacetime geometry, but for now we present an experimental proposal that could, in principle, serve such a purpose. In the following, we restore the units.    

As we have seen, the metric of Melvin's electrical universe in stationary cylindrical coordinates reads
\begin{equation}\diff s^{2}=\lambda^{2}(\rho)[-c^{2}\diff t^{2}+\diff \rho^{2}+\diff z^{2}]+\frac{\rho^{2}}{\lambda^{2}(\rho)}\diff \varphi^{2}, \end{equation} 
with $\lambda(\rho)=1+\frac{\rho2}{L^{2}}$, where $L$ is the characteristic length scale for Melvin's electric universe given by 
\begin{equation}
L=\frac{2c^{2}}{EG^{1/2}}\sim \bar{E}^{-1}\times 10^{24}cm,   
\end{equation}
$\bar{E}$ being a numerical value of the electric field in Gaussian CGS units (base units $cm^{-1/2}g^{1/2}s^{-1}$). It is not clear, on the empirical basis, what value should we asignd to the background electric field. As a reference value we may take $\bar{E}\sim 10^{8}$ \cite{Pulsar}, and thus $L\sim 10^{16}cm$. 

Consider now a ring-shaped Sagnac interferometer of radius $R$, centered around the symmetry axis of Melvin's electric universe. Let the ring rotate with constant angular velocity $\Omega_{\text{Ring}}>0$. In non-inertial coordinates $t'=t$, $\rho'=\rho$, $z'=z$ and $\varphi'=\varphi-\Omega_{\text{Ring}}t$ adapted to the rotating frame of the ring, Melvin's metric becomes
\begin{align}
\diff s^{2}=&\lambda(\rho)[-c^{2}\diff t^{2}+\diff \rho^{2}+\diff z^{2}]+\frac{\rho^{2}}{\lambda^{2}(\rho)}(\diff \varphi'+\Omega_{\text{Ring}}\diff t)^{2}\nonumber\\
=&\left(-c^{2}\lambda^{2}(\rho)+\frac{\rho^{2}\Omega^{2}_{\text{Ring}}}{\lambda^{2}(\rho)}\right)\diff t^{2}+\lambda^{2}(\rho)[\diff \rho^{2}+\diff z^{2}]\nonumber\\
&+\frac{\rho^{2}}{\lambda^{2}(\rho)}\diff\varphi'^{2}+\frac{\rho^{2}\Omega_{\text{Ring}}}{\lambda^{2}(\rho)}(\diff\varphi' \diff t+\diff t\diff\varphi').
\end{align}
In the Sagnac interferometric experiment, we are interested in what happens with beams counter-propagating along the rotating ring. So we can set $\rho=R$ and $z=0$ and focus on Melvin's metric at fixed radial distance in the horizontal plane,
\begin{align}
&\diff s_{\text{Ring}}^{2}
=\left(-c^{2}\lambda^{2}(R)+\frac{R^{2}\Omega^{2}_{\text{Ring}}}{\lambda^{2}(R)}\right)\diff t^{2}\nonumber\\
&+\frac{R^{2}}{\lambda^{2}(R)}\diff\varphi'^{2}+\frac{R^{2}\Omega_{\text{Ring}}}{\lambda^{2}(R)}(\diff\varphi' \diff t+\diff t\diff\varphi')\\&\approx -\lambda^{2}(R)c^{2}\diff t^{2}+\frac{R^{2}}{\lambda^{2}(R)}\diff\varphi'^{2}+\frac{R^{2}\Omega_{\text{Ring}}}{\lambda^{2}(R)}(\diff\varphi' \diff t+\diff t\diff\varphi'),\nonumber 
\end{align}
where the approximation holds for $\Omega_{\text{Ring}}R\ll c$. Since Melvin's scale $L\sim 10^{18}cm$ is very large, for realistic values of $R$ we have $\lambda(R)\approx 1$; Melvin's universe becomes increasingly flat as we approach the symmetry axis. 

\begin{figure}[h!]
    \centering
    \includegraphics[scale=0.8]{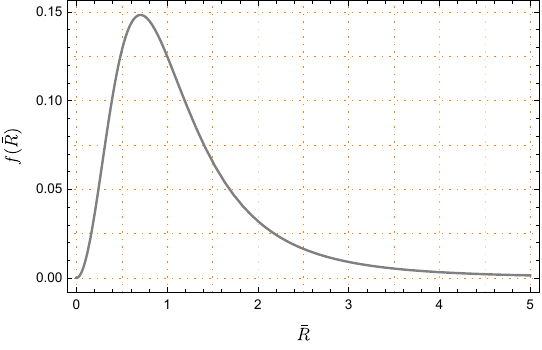}
    \caption{Plot of $f(\bar{R})$. For our analysis, the relevant range of parameters is $\bar{R}\ll1$.}
    \label{fig}
\end{figure}

This was in the case of a rotating ring. However, comparing with the effective NC metric (\ref{NCmetric}) we see, up to the first order in $\theta$, that even if the ring was \textit{stationary}, the counter-propageting beams of \textit{charged} matter (scalars or spinors) in the NC deformed Melvin's electric universe behave \textit{as if} the ring was rotating with angular velocity $\Omega_{\text{Ring}}=-\Omega_{\text{NC}}$; since $\Omega_{\text{NC}}=-\frac{1}{2}qE\kbar/\hbar\sim \kbar$, it is consistent to ignore the term $\Omega_{\text{Ring}}^{2}$. Therefore, if we set up a stationary Sagnac ring interferometer around the symmetry axis of Melvin's electric universe, and use a beam splitter to branch out an electrically charged quantum particle around the ring (with equal absolute velocity with respect to the ring), due to spacetime noncommutativity, the observed Sagnac phase shift would be the same as if the ring interferometer itself were rotating with angular velocity $\Omega_{\text{Ring}}=\frac{1}{2}qE\kbar/\hbar$. Therefore, the metric on which we base our Sagnac shift calculation is
\begin{align}
\diff s^{2}_{\text{Ring}}=&-\lambda^{2}(R)c^{2}\diff t^{2}+\frac{R^{2}}{\lambda^{2}(R)}\diff \varphi'^{2}\nonumber\\
&-\frac{R^{2}\Omega_{\text{NC}}}{\lambda^{2}(R)}(\diff t\diff \varphi'+\diff\varphi'  \diff t).
\end{align}
Following the general recipe from \cite{Sagnac2}, we start with the $4$-velocity field for the time-like congruence of rotating observers that constitute the ring, with components 
\begin{align}
u^{0}&=\frac{1}{\sqrt{-g_{00}}}=\frac{1}{c\lambda(R)},\\ u_{0}&=-\sqrt{-g_{00}}=-c\lambda(R),\\ u_{\varphi'}&=g_{\varphi'0}u^{0}=-\frac{R^{2}\Omega_{\text{NC}}}{c\lambda^{3}(R)}, 
\end{align}
and the gravito-magnetic vector potential
\begin{equation}
A_{\varphi'}^{G}=c^{2}\frac{u_{\varphi'}}{u_{0}}=\frac{R^{2}\Omega_{\text{NC}}}{\lambda^{4}(R)}.    \end{equation} 
We calculate the Sagnac phase shift as a gravito-magnetic Aharonov-Bohm phase shift,
\begin{align}
\Delta\phi&=\frac{2mu_{0}}{c\hbar}\oint_{\text{Ring}}\boldsymbol{A}^{G}\cdot\diff\boldsymbol{x}
=-\frac{2m\sqrt{-g_{00}}}{c\hbar}\int_{0}^{2\pi}A_{\varphi'}^{G}\diff\varphi'\nonumber\\
&=-\frac{4mc^{4}\Omega_{\text{NC}}}{\hbar G E^{2}}f(\bar{R}), 
\end{align}
where we introduced the geometrical factor,
\begin{equation}
f(\bar{R})=\frac{\bar{R}^{2}}{\left(1+\bar{R}^{2}\right)^{3}} \end{equation}
with dimensionless parameter $\bar{R}=R/L$. The function $f(\bar{R})$ is given in Figure \ref{fig}.
With the NC angular speed given by $\Omega_{\text{NC}}=-\frac{1}{2}qE\kbar/\hbar$, the NC Sagnac phase shift can be expressed as
\begin{equation}
\Delta\phi=\frac{4\pi q \kbar}{El^{2}_{\text{Planck}}\lambda_{\text{Compton}}}f(\bar{R}). \end{equation}
For a particle of mass $m$ and charge $q$, after setting the magnitude of the background electric field $E$, the value of the NC Sagnac phase shift depends solely on the radius of the interferometer through the geometric factor $f(\bar{R})$. For a small radius, i.e., for $\bar{R}=R/L\ll 1$, we have $f(\bar{R})\sim\bar{R}^{2}$, and the shift comes down to
\begin{equation}
\Delta\phi\sim-\frac{4m}{\hbar}\pi R^{2}\Omega_{\text{NC}}=\frac{4m}{\hbar}A\Omega_{\text{Ring}},    
\end{equation}
which is exactly the result for flat Minkowski spacetime ignoring the $\mathcal{O}(\Omega^{2}R^{2}/c^{2})$ terms (the relativistic corrections). More explicitly, near the symmetry axis Melvin's electric universe, where the spacetime is approximately flat, the (non-relativistic) NC shift is 
\begin{equation}
\Delta\phi=\frac{2m q \Phi_{E}\kbar}{\hbar^{2}},
\end{equation}
where $\Phi_{E}=\pi R^{2} E$ is the electric flux through the ring interferometer. For an electron and the reference value of the background electric field, we have
\begin{equation}
\Delta\phi_{\text{electron}}\approx  8\times 10^{26}\;\mathrm{cm}^{-3}\times R^{2}\kbar.    
\end{equation}
Suppose that $\kbar\sim l_{\text{Planck}}\sim 10^{-38}\;\mathrm{cm}$. For a typical phase shift resolution of $10^{-3}\;\mathrm{rad}$, to observe the NC Sagnac effect, we need at least $R\sim 10^{4}\;\mathrm{cm}$ or larger. But if we expect to find the spacetime noncommutativity at some larger scale then $l_{\text{Planck}}$, we can use a smaller interferometer. With a tabletop size of $R\sim 1 \;\mathrm{cm}$, we should be able to detect the NC Sagnac shift for $\kbar\sim 10^{-30}\;\mathrm{cm}$ or larger. We can therefore say that the NC Sagnac effect can serve as a efficient probe of the small-scale structure of spacetime.     

\section{Discussion}

We have shown that dynamical effects of NC geometry on charged matter fields can be interpreted in a purely kinematical fashion as a transition to a rotating frame of reference with angular speed determined by the noncommutativity scale and the electric charge of the field. There are several points to be made here. First, the conclusion holds only in a perturbative regime, at least up to the first order in $\kbar$. It would be interesting to see whether the interpretation holds in higher orders, especially given that the angular speed appears at most in its quadratic power in rotating coordinates, while $\kbar$-corrections can, in general, be of any order. Second, Melvin's electric universe is not a very realistic model; it was a convenient choice for implementing the main idea. To make some empirically relevant theoretical predictions, we would need to conduct a similar analysis
in a more realistic cosmological spacetime, but with enough symmetry to allow Killing twists. Third, there is a very general issue about ``choosing a twist'' that we have explained in the introduction. In particular, we can only make our case employing a Killing twist, because this allows us to keep the geometry undeformed. Although this seems fairly reasonable, what happens in a general NC theory remains an open question. Finally, there is a technical issue in the proposed Sagnac setup. Namely, the background electric field pushes the charged particle, which propagates around the ring interferometer, in the $z$-direction perpendicular to the ring. To enforce the constraint of moving around the ring, we have to introduce an infinite confining potential that would disrupt the external electric field. This setback can be avoided by promoting the ring interferometer to a cylindrical one, thus allowing free propagation in the $z$-direction. As far as the Sagnac interference experiment is concerned, this relaxation does not change the resulting effect. As an outlook, we propose a systematic research of the relation between NC twist deformations and reference frame transformations exemplified in this paper. It seems significant that dynamical NC effects can be interpreted kinematically as some kind of inertial force.  

\section*{Acknowledgements}
We want to thank Marija Dimitrijevi\'{c} \'{C}iri\'{c} for useful discussions. The work of D.D. and D.G. is supported by the funding provided by the Faculty of Physics, University of Belgrade, through grant number 451-03-136/2025-03/200162 by the Ministry of Science, Technological Development and Innovations of the Republic of Serbia. The research was supported by the Science Fund of the Republic of Serbia, grant number TF C1389-YF, Towards a Holographic Description of Noncommutative Spacetime: Insights from Chern-Simons Gravity, Black Holes and Quantum Information Theory - HINT.

\appendix

\onecolumngrid

\section{ Abelian twist}

We follow \cite{LA}. Let $M$ be a smooth manifold, and let $\Xi=\Gamma(TM)$ be the Lie algebra of smooth vector fields on $M$.  Denote by $U\Xi$ its universal enveloping algebra (so that for $X,Y\in\Xi$ the commutator $XY-YX$ is identified with the Lie bracket $[X,Y]$). A \emph{twist} $\mathcal{F}\in U\Xi\otimes U\Xi$ is an \emph{invertible} element that satisfies the usual Drinfeld‐twist axioms for deforming a Hopf algebra. In particular, one may consider the class of \emph{abelian} twists of the form
\begin{align}
  \mathcal{F}=
  \exp\Bigl(-\tfrac{i}{2}\theta^{IJ}X_I\otimes X_J\Bigr),
    \label{eq:abeliantwist}
\end{align}
where $X_I$ $(I=1,\dots,s\leq D)$ are \emph{mutually commuting} vector fields, i.e. $[X_I,X_J]=0$, and $\theta^{IJ}$ a constant antysymmetric matrix (note that the number of vector fields $s$ need not be equal to the total number of spacetime dimensions $D$; in that case the vectors defining the twist span a
subspace of the $D$-dimensional tangent space at each point). In expanded form, the inverse of the twist is given by
\begin{equation}
  \mathcal{F}^{-1}
  =\exp\Bigl(\tfrac{i}{2}\theta^{IJ}X_I\otimes X_J\Bigr)=
  1\otimes1
  +\tfrac{i}{2}\theta^{IJ}X_I\otimes X_J
  -\tfrac{1}{8}\theta^{I_1J_1}\theta^{I_2J_2}
    X_{I_1}X_{I_2}\otimes X_{J_1}X_{J_2}
  +\cdots,
\end{equation}
It is convenient to introduce the shorthand notation (the summation over the multi‐index $\alpha$ is implied) 
\begin{equation}
  \mathcal{F}^{-1}  \equiv
  \bar f^{\alpha}\otimes\bar f_{\alpha}.
\end{equation}
Let now \(A = C^\infty(M)\) be the algebra of smooth functions on the manifold \(M\). Given a twist \(\mathcal{F}\), we deform \(A\) into the corresponding NC algebra \(A_\star\) by defining the new product,
\[
  f\star g = \bar f^{\alpha}(f)\bar f_{\alpha}(h)=\mu\!\Bigl(\mathcal{F}^{-1}
    (f\otimes g)\Bigr),
\]
where $\mu(f\otimes g)=f\cdot g$ is the ordinary commutative pointwise multiplication. Since the vector fields are, in general, $x$-dependent, so is the $\star$-product. If the fields \(X_I\) commute, the \(\star\)-product is \textit{associative}. This geometric definition of a $\star$-product is a direct generalization of the standard Moyal–Weyl-Groenewold \(\star\)-product for which (assuming that there exists a global coordinate system on the manifold) $X_I=\delta^{\mu}_{I}\partial_\mu$ and 
\begin{equation}
(f\star g)(x)=\mu\left(e^{\tfrac{i}{2}\theta^{\mu\nu}\partial_{\mu}\otimes\partial_{\nu}}(f\otimes g)\right)=f(x)g(x)+\frac{i}{2}\theta^{\mu\nu}\partial_{\mu}f(x)\partial_{\nu}g(x)+\mathcal{O}(\theta^{2}).
\end{equation}

\section{Fermions}

Here we show that the same effect of NC rotation can be obtained for fermionic matter. Classical equations of motion for a massive Dirac spinor $\psi$ in our signature are given as 
\begin{equation}\label{Dirac}
(\slashed{\mathcal{D}}-m)\psi=0\,,
\end{equation}
where $\mathcal{D}_\mu=\partial_\mu+\frac{1}{4}\omega^{ab}_\mu\gamma_{ab}-iqA_\mu$, $\psi=(\psi_1,\psi_2,\psi_3,\psi_4)^{\mathrm{T}}$ and $\slashed{\mathcal{D}}=\gamma^\mu\mathcal{D}_\mu$. Gamma matrices are defined as 
\begin{equation}
\gamma^0=\left(
\begin{array}{cccc}
 0 & 0 & 1 & 0 \\
 0 & 0 & 0 & 1 \\
 1 & 0 & 0 & 0 \\
 0 & 1 & 0 & 0 \\
\end{array}
\right) \hspace{5mm}\gamma^1=\left(
\begin{array}{cccc}
 0 & 0 & 1 & 0 \\
 0 & 0 & 0 & -1 \\
 -1 & 0 & 0 & 0 \\
 0 & 1 & 0 & 0 \\
\end{array}
\right)
\end{equation}
\begin{equation}
    \gamma^2=\left(
\begin{array}{cccc}
 0 & 0 & 0 & 1 \\
 0 & 0 & 1 & 0 \\
 0 & -1 & 0 & 0 \\
 -1 & 0 & 0 & 0 \\
\end{array}
\right)\hspace{5mm}\gamma^3=\left(
\begin{array}{cccc}
 0 & 0 & 0 & -i \\
 0 & 0 & i & 0 \\
 0 & i & 0 & 0 \\
 -i & 0 & 0 & 0 \\
\end{array}
\right)
\end{equation}
while $\gamma_{ab}=\frac{1}{2}[\gamma_a,\gamma_b]$.
The NC corrections to these equations are introduced analogously as before, following \cite{K2}. The NC action with abelian killing twist is  
\begin{equation}
\int \diff^4 x \;|e|\star\hat{\psi}\star \left(\slashed{\partial}\hat{\psi}  +\frac{1}{4}\slashed{\omega}\star\hat{\psi} -iq\slashed{A} \star\hat{\psi}-m\hat{\psi} \right),
\end{equation}
where again we can drop the $\star$-product when multiplying the vielbein or the spin-connection components. 
To perform an explicit computation, we consider the case without a cosmological constant. 
We choose diagonal vielbeins as 
\begin{equation}
 e^a_{\;\;\mu}=   \left(
\begin{array}{cccc}
 \lambda(\rho) & 0 & 0 & 0 \\
 0 & \lambda(\rho) & 0 & 0 \\
 0 & 0 & \lambda(\rho) & 0 \\
 0 & 0 & 0 & \frac{\rho}{\lambda(\rho)} \\
\end{array}
\right),
\end{equation}
and the inverse vielbein is 
\begin{equation}
     e_a^{\;\;\mu}=\left(
\begin{array}{cccc}
 \frac{1}{\lambda(\rho)} & 0 & 0 &0 \\
 0 & \frac{1}{\lambda(\rho)} & 0 & 0 \\
 0 & 0 & \frac{1}{\lambda(\rho)} & 0 \\
 0 & 0 & 0 & \frac{\lambda(\rho)}{\rho} \\
\end{array}
\right)\,.
\end{equation}
As we are in the scope of general relativity, the torsion $T^a_{\mu\nu}=\partial_\mu e^a_{\;\;\nu}-\partial_\nu  e^a_{\;\;\mu}+\omega_{\mu \;b}^{a} e^b_{\;\;\nu}- \omega_{\nu\;b}^{a} e^b_{\;\;\mu}$ is set to zero, and from this condition we can read off the spin-connection components,
\begin{align}
\omega_{\varphi}^{13}=-\omega_{\varphi}^{31}=
\frac{2-\lambda(\rho)}{\lambda^3(\rho)},\nonumber\\ \label{spink}
\omega_z^{12}=-\omega_z^{21}=\omega_t^{10}=-\omega_t^{10}=\frac{\rho E^2}{2\lambda(\rho)}.
\end{align}
It can be shown \cite{K2} that the first-order NC effects include the following NC term, 
\begin{equation}\label{NCF}
-\frac{\kbar}{2}qF_{tz}\gamma^t\partial_\varphi \psi=\frac{\kbar}{2\lambda(\rho)} qE\gamma^0\partial_\phi\psi,
\end{equation}
on the left-hand side of the equation
\eqref{Dirac}. Importantly, this correction can be obtained by working with the commutative equation for the effective vielbein
\begin{equation}
 \Tilde{e}^a_{\;\;\mu}=   \left(
\begin{array}{cccc}
 \lambda(\rho) & 0 & 0 & 0 \\
 0 & \lambda(\rho) & 0 & 0 \\
 0 & 0 & \lambda(\rho) & 0 \\
 \frac{-q E \kbar \rho }{2\lambda(\rho)} & 0 & 0 & \frac{\rho}{\lambda(\rho)} \\
\end{array}
\right),
\end{equation}
The inverse vielbein (up to the first order in $\kbar$) is given by 
\begin{equation}
     \Tilde{e}_a^{\;\;\mu}=\left(
\begin{array}{cccc}
 \frac{1}{\lambda(\rho)} & 0 & 0 & \frac{ q E \kbar}{2\lambda(\rho)} \\
 0 & \frac{1}{\lambda(\rho)} & 0 & 0 \\
 0 & 0 & \frac{1}{\lambda(\rho)} & 0 \\
0 & 0 & 0 & \frac{\lambda(\rho)}{\rho}\\
\end{array}
\right).
\end{equation}
For this effective NC metric, spin-connection components are the same as in \eqref{spink}, with the additional nonzero component 
\begin{equation}
    \omega_t^{13}=-\omega_t^{31}=-\frac{qE\kbar(2-\lambda(\rho))}{4\lambda^3(\rho)}.
\end{equation}
The Dirac equation for the effective NC metric  is given by the following set of four coupled partial differential equations,
\begin{align}\nonumber
\frac{\left(\partial_\rho\psi_3+\partial_t\psi_3+\partial_z\psi_4\right)}{\lambda(\rho)}-\frac{i \lambda(\rho)
   \partial_\phi \psi_4}{ \rho }+\frac{ \left(3E^2\rho^2-8iEq\rho z\lambda(\rho)+4\right)\psi_3}{8\rho  \lambda^{2}(\rho)}-m\psi_1+\frac{ \kbar E q }{2\lambda(\rho)}\partial_\phi\psi_3=0\,,\\\nonumber
   \frac{ \left(\partial_z\psi_3-\partial_\rho\psi_4+\partial_t\psi_4\right)}{\lambda(\rho)}+\frac{i \lambda(\rho)
   \partial_\phi\psi_3}{ \rho }+\frac{ \left(-3 E^2 \rho ^2-8 i E q \rho  z \lambda(\rho)-4\right) \psi_4}{8\rho \lambda^{2}(\rho)}-m \psi_2+\frac{ \kbar E q }{2\lambda(\rho)}\partial_\phi \psi_4=0\,,\\\nonumber
   \frac{\left(-\partial_z\psi_2-\partial_\rho\psi_1+\partial_t\psi_1\right)}{\lambda(\rho)}+\frac{i \lambda(\rho)
   \partial_\phi\psi_2}{ \rho }+\frac{ \left(-3 E^2 \rho ^2-8 i E q \rho  z \lambda(\rho)-4\right) \psi_1}{8\rho  \lambda^2(\rho)}-m \psi_3+\frac{ \kbar E q }{2\lambda(\rho)}\partial_\phi \psi_1=0\,,\\\nonumber
   \frac{ \left(-\partial_z\psi_1+\partial_\rho\psi_2+\partial_t\psi_2\right)}{\lambda(\rho)}-\frac{i \lambda(\rho)
   \partial_\phi\psi_1}{ \rho }+\frac{\left(3E^2\rho^2-8iEq\rho z\lambda(\rho)+4\right) \psi_2}{8\rho  \lambda^{2}(\rho)}-m \psi_4+\frac{ \kbar E q }{2\lambda(\rho)}\partial_\phi \psi_2=0\,.
\end{align}
It is then easy to see that the NC corrections are precisely of the form \eqref{NCF}. The effective metric obtained from the effective vielbein is given by \eqref{NCmetric}. A similar conclusion can be made for $\Lambda>0$.

\end{document}